\documentclass[epj]{svjour}
\usepackage{graphics}
\begin{document}
\title{The Goldberger-Miyazawa-Oehme sum rule revisited}
\author{V.V. Abaev\inst{1}, P. Mets\"a\inst{2}, and M.E. Sainio\inst{3}
}                     
%
%
\institute{Petersburg Nuclear Physics Institute, Gatchina 188300, Russia \and  Department of Physical Sciences, P.O. Box 64, 
 00014 University of Helsinki, Finland 
\and Helsinki Institute of Physics, P.O. Box 64, 00014 University of Helsinki, Finland 
          }
\date{Received: date / Revised version: date}
%
\abstract{
The Goldberger-Miyazawa-Oehme sum rule is used to extract the
pion-nucleon coupling constant from experimental $\pi$N information.
Chiral perturbation theory is exploited in relating the pionic
hydrogen $s$-wave level shift and width results to the appropriate
scattering lengths. The deduced value for the coupling is $f^2 = 0.075 \pm 0.002$,
where the largest source of uncertainty is the determination of
the $s$-wave $\pi^- p$ scattering length from the atomic level shift measurement.
} 
\maketitle
\section{Introduction}
\label{intro}

The Goldberger-Miyazawa-Oehme \cite{gmo} sum rule relates the coupling constant
of the pion-nucleon interaction, $f^2$, to the difference of the $\pi^-p$ and $\pi^+p$
$s$-wave scattering lengths and to the weighted integral of the difference of
the $\pi^-p$ and $\pi^+p$ total cross sections. By evaluating the forward dispersion relations
at the physical threshold, $\omega=\mu$, we obtain \cite{hw}
\begin{eqnarray}
(1+\mu/m)(a_{\pi^-p} - a_{\pi^+p})/\mu=4 f^2/(\mu^2-\omega_B^2) + 2 \, J^-, 
\end{eqnarray}
where $\mu$ is the (charged) pion mass, $m$ is the proton mass, $\omega_B=-\mu^2/2m$
and
\begin{eqnarray}
J^- =   1/(4\pi^2) \; \int^\infty_0 (\sigma^{\rm Tot}_{\pi^-p}-\sigma^{\rm Tot}_{\pi^+p})/\omega \; dk.
\end{eqnarray}
The $s$-wave $\pi^-p$ and $\pi^+p$ scattering lengths are denoted
by $a_{\pi^-p}$ and $a_{\pi^+p}$ respectively and $\omega=\sqrt{\mu^2+k^2}$, where
$k$ is the pion laboratory momentum.

It is well known that eq. (1) is not a precise means to
determine the value of the pion-nucleon coupling $f^2$, but
it is the natural first step in the process of extracting the value from 
the pion-nucleon data. Furthermore, the sum rule makes the connection between the 
uncertainties in the data and in the coupling constant more transparent.

It is the aim of the present paper to examine the ingredients of eq. (1), two scattering lengths 
and integral $J^-$, and determine the pion-nucleon coupling strength $f^2$. 
The implications of the pionic hydrogen measurements to the sum rule are discussed in
sect. 2, the $s$-wave $\pi^+p$ scattering length  is addressed in sect.
3 and the integral $J^-$ in sect. 4. Section 5 combines the information to
extract a value for $f^2$ and in sect. 6 the conclusions are drawn.  

\section{Pionic hydrogen} 
\label{hydrogen}

The accurate measurements of the properties of pionic hydrogen at PSI
provide a source of information on the pion-nucleon amplitude very close to the physical 
threshold. The strong interaction 
level shift $\epsilon_{1s}$, determined through the difference
of the electromagnetic and measured transition energy difference
\begin{eqnarray}
\epsilon_{1s}=E^{\rm em}_{3p-1s} - E^{\rm meas}_{3p-1s},
\end{eqnarray}
has been measured to a precision of about 0.2 \% \cite{simons}
\begin{eqnarray}
\epsilon_{1s} = -7.120 \pm 0.008 \pm 0.009 \; \;{\rm eV}.
\end{eqnarray}
The first error is due to statistics and the second due to systematics.
There is a considerable improvement in accuracy compared with the earlier 
result \cite{schroder} 
\begin{eqnarray}
\epsilon_{1s} = -7.108 \pm 0.013 \pm 0.034 \; \;{\rm eV},
\end{eqnarray}
mainly because the molecular effects are better under control.
However, the improvement is not reflected in the extraction
of the $\pi^-p$  $s$-wave scattering length. The connection between 
the level shift and the scattering lengths is provided
by the formula \cite{gasser}
\begin{eqnarray}
\epsilon_{1s}=-2 \alpha^3 \,\mu^2_c (a^+_{0+}+a^-_{0+}) (1+\delta_\epsilon),
\end{eqnarray}
where $\mu_c$ is the reduced mass of the $\pi^-p$ system, $\alpha \simeq 1/137.036$ 
is the fine structure constant and $a^+_{0+}$, $a^-_{0+}$ are the
isoscalar and isovector $s$-wave $\pi$N scattering lengths respectively.
The quantity $\delta_\epsilon$ evaluated next-to-leading order
in isospin breaking and in the low-energy expansion has
the value \cite{gasser}
\begin{eqnarray}
\delta_\epsilon = (-7.2 \pm 2.9) \times 10^{-2}.
\end{eqnarray}
The uncertainty of the $\pi^-p$ scattering length determined
with eq. (6) is then
dominated by the uncertainty in $\delta_\epsilon$, which in turn is 
dominated by the largely unknown low-energy constant $f_1$ \cite{gasser}.
Potential models produce typically smaller correction factors $\delta_\epsilon$
with considerably smaller uncertainties \cite{sigg,elw}. However, here
we rely on the chiral perturbation theory result \cite{gasser} for $\delta_\epsilon$
and consequently the measured level shift of eq. (4) gives
\begin{eqnarray}
a_{\pi^-p} = 0.0933 \pm 0.0029 \; \; 1/\mu,
\end{eqnarray}
if errors in eq. (4) are added linearly.
Here the identification $a_{\pi^-p} = a^+_{0+} + a^-_{0+}$ has been made.

The Deser-type formula \cite{deser} relating the isovector $s$-wave scattering 
length $a^-_{0+}$ to the level width is \cite{sigg}
\begin{eqnarray}
\Gamma_{1s}=8 \alpha^3 \, \mu^2_c \,q_0 (1+1/P)[a^-_{0+}(1+\delta_\Gamma)]^2,
\end{eqnarray}
where $q_0$ is the centre-of-mass momentum of the $\pi^0$ in the charge
exchange reaction and $P=1.546\pm0.009$ is the Panofsky ratio \cite{spuller},
{\it i.e.} the ratio of the cross sections 
$\sigma(\pi^-p\rightarrow \pi^0n)/\sigma(\pi^-p\rightarrow \gamma n)$ at the 
threshold. The correction factor $\delta_\Gamma$ has been evaluated in
leading order in chiral perturbation theory with the result \cite{zemp}
\begin{eqnarray}
\delta_\Gamma = (0.6 \pm 0.2) \times 10^{-2}.
\end{eqnarray}
A potential model calculation \cite{sigg} would give a different
sign, but this correction factor is rather small in any case.
The factor $\delta_\Gamma$ depends on the low-energy
constant $f_2$ and, therefore, the uncertainty is smaller than
for the factor $\delta_\epsilon$, which depends on the constant $f_1$.

The $s$-wave isovector scattering length can be solved from eq. (9)
\begin{eqnarray}
a^-_{0+}=\frac{0.08933}{1+\delta_\Gamma} \, \sqrt{\frac{\Gamma_{1s}}{0.868 \; {\rm eV}}} \; \;1/\mu.
\end{eqnarray}
The measured result for the width \cite{schroder}
$\Gamma_{1s}=0.868 \pm0.040\pm0.038 \; \; {\rm eV}$
then gives for the scattering length
$a^-_{0+}=0.0888 \pm 0.0040 \; 1/\mu$. The relatively large error bar
is largely due to the uncertainty in estimating the Doppler
broadening. The new preliminary result for the width \cite{simons2}
with improvement in this respect, $\Gamma_{1s} = 0.823 \pm 0.019 \; {\rm eV}$,
would give $a^-_{0+} = 0.0865 \pm 0.0010 \; 1/\mu$.

\section{Discrete phase shift analysis for $\pi^+p$}
\label{pin}

For the $\pi^+p$ interaction a discrete phase shift analysis has been performed
in the range $k=0.077-0.725$ GeV/c at 77 different momenta \cite{abaev}. 
Tromborg corrections \cite{tromborg} have been used to extract the hadronic amplitudes
from the experimental data. Forward scattering constraints \cite{pekko} have been applied
iteratively in the analysis. Additional constraints from total 
inelastic cross sections were used for partial wave inelasticities.
Special care has been taken to incorporate in the analysis the experimental resolutions and acceptances. 
In this way only a small number of points had to be eliminated due to inappropriate 
angular dependence. The normalizations were allowed to float according to experimental
information on the systematic uncertainty.
Furthermore, the behaviour of the zero trajectories for transversity amplitudes was monitored
to keep a smooth variation in the shape of observables and the extrapolation
to the physical threshold was stabilized in this manner. 
Figure 1 shows the phase shift $\delta_{S31}$ close to the threshold normalized with the product of the 
absolute value of the $s$-wave scattering 
length $a_{S31}$ and the centre-of-mass momentum $q$. 
The result can be parametrized with analytical forms in three different
$q^2$ ranges such that in the lowest energy range the effective range
approximation is used. The two dividing momenta have been taken as free
parameters. It turns out that a two-parameter effective range
approximation is valid up to $q \simeq 0.14$ GeV/c.
The result for the scattering length $a_{S31}$
\begin{eqnarray}
a_{S31} \equiv a_{\pi^+p}= -0.0764 \pm 0.0014 \; \; 1/\mu
\end{eqnarray}
is very stable.
This result completes the discussion of the left hand side of eq. (1).
The result can be compared with the recent figure of Matsinos {\it et al.}
\cite{matsinos} $a_{\pi^+p}=-0.0751 \pm 0.0039 \; 1/\mu$ based on a fit with a low-energy
model and with the FA02 solution of the GWU/VPI group \cite{arndt} 
$a_{\pi^+p}=-0.0911 \pm 0.0014 \; 1/\mu$. For the KH80 solution
\cite{koch} of the Karlsruhe group  $a_{\pi^+p}=-0.1010 \pm 0.0040 \; 1/\mu$.
\begin{figure}
\resizebox{0.48\textwidth}{!}
{\includegraphics{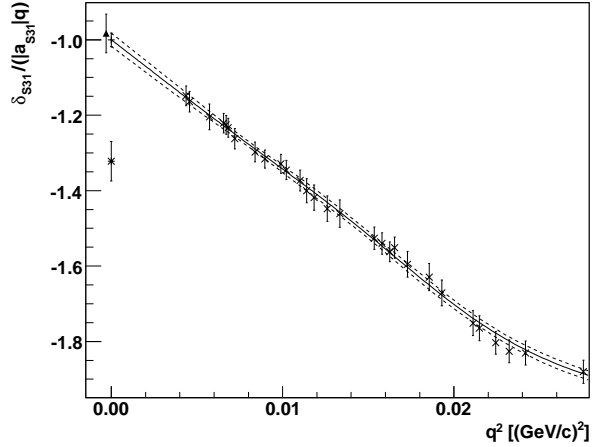}}
\caption{The $\pi^+p$ $s$-wave phase shift $\delta_{S31}$ normalized with $|a_{S31}| \, q$.
The asterix denotes the KH80 value. The result of Matsinos {\it et al.} is shown by the triangle.}
\label{fig:1}
\end{figure}

\section{Integral $J^-$}
\label{int}

The integral $J^-$ contains input from a number of  sources.
Results from total cross section measurements are available in the range
0.16-640 GeV/c for the $\pi^-p$ scattering and 0.16-340 GeV/c for the
$\pi^+p$ scattering \cite{durham}. These data have been corrected for the electromagnetic
effects up to $k=0.725$ GeV/c using the formalism of Tromborg \cite{tromborg}.
The corrections have been published only up to 0.655 GeV/c, but here
we employ a smooth extrapolation up to 0.725 GeV/c. The question of electromagnetic
corrections above the Tromborg range has been discussed in a number of
articles \cite{rix,bc}. Here we do not, however,  correct the data above
the limit 0.725 GeV/c, but we estimate the corresponding uncertainty in the value
of $J^-$ by making use of the prescription in \cite{elt}. A smooth curve through the
experimental points is drawn using the expansion technique \cite{hohler}, which
guarantees that forward dispersion relations are satisfied. Details can be found
elsewhere \cite{pekko}.

The integral $J^-$ converges relatively slowly and the high-energy behaviour
of the cross sections $\sigma_{\pi^\pm p}^{\rm Tot}$ plays a significant role.
Here we consider three different Regge type parametrizations for the asymptotic forms, {\it i.e.}
forms which give estimates of the total cross sections beyond 350 GeV/c:
the parametrizations of H\"ohler \cite{hohler}, Donnachie and Landshoff \cite{dl}
and Gauron and Nicolescu \cite{gn}. In the range 10 - 350 GeV/c the recent parametrization 
of the Particle Data Group \cite{PDG} has been employed in addition to the Regge
forms quoted above and the result from the fit with the expansion technique \cite{pekko}.
The results of these parametrizations have been summarized in table 1.
The displayed number of significant figures is there only to help
in the addition of the contributions of the different momentum ranges.
\begin{table}
\caption{ Contributions to $J^-$ (mb) of the different high-energy ranges of the laboratory momentum
$k$.}
\label{tab:1}       
\begin{center}
\begin{tabular}{lll}
\hline\noalign{\smallskip}
Input & 10-350 GeV/c & 350- GeV/c  \\
\noalign{\smallskip}\hline\noalign{\smallskip}
H\"ohler \cite{hohler} & 0.08786 & 0.01787 \\
Donnachie-Landshoff \cite{dl} & 0.09968 & 0.02514 \\
Gauron-Nicolescu \cite{gn} & 0.10665 & 0.02012 \\
PDG \cite{PDG} & 0.09587 & - \\
Present work  & 0.09609 & - \\
\noalign{\smallskip}\hline
\end{tabular}
\end{center}
\end{table}
The momentum range from the threshold to 10 GeV/c is split into two at 2.03 GeV/c.
For the low-energy section from the threshold to 2.03 GeV/c the experiment covers
only part of the range. From the discussion in sects. 2 and 3 we can, 
by invoking isospin invariance, deduce values 
for the $\pi^\pm p$ total cross sections at the physical threshold with the results
$\sigma^{\rm Tot}_{\pi^-p} =5.80 \pm 0.27 \;{\rm mb}$ and 
$\sigma^{\rm Tot}_{\pi^+p} = 1.47 \pm 0.05 \;{\rm mb}$.
In the range from 0.08 to 0.16 GeV/c there is information
on the angular distributions, but no modern total cross section data. For this
range we have performed a phase shift analysis with elastic $\pi^\pm p$ input and
fixed-$t$ constraints. Also, results from the discrete phase shift analysis for
$\pi^+ p$ discussed in sect. 3 have been used as cross checks. 
The remaining piece from the threshold to 0.08 GeV/c has been covered
by a smooth interpolation. The contributions to integral $J^-$ can then be evaluated with
the results displayed in table 2. There, in addition, results from the phase shift
analyses of the Karlsruhe group and the GWU-VPI group are shown.
\begin{table}
\caption{ Contributions to $J^-$ (mb) of the low and intermediate energy ranges of the laboratory momentum
$k$.}
\label{tab:2}       
\begin{center}
\begin{tabular}{lll}
\hline\noalign{\smallskip}
Input & 0-2.03 GeV/c & 2.03-10 GeV/c  \\
\noalign{\smallskip}\hline\noalign{\smallskip}
KH80 \cite{koch} & -1.27853 & 0.10691 \\
KA84 \cite{koch2} & -1.31266 & 0.13802 \\
FA02 \cite{arndt} & -1.30213 & - \\
Present work  & -1.29757 & 0.12046 \\
\noalign{\smallskip}\hline
\end{tabular}
\end{center}
\end{table}
A detailed discussion of the contributions of different momentum ranges can be found, 
{\it e.g.}, in ref. \cite{elt}.

Combining the values from different momentum 
slices yields the value for $J^-$ given in table 3. Also, results from some earlier
evaluations have been displayed there. 
\begin{table}
\caption{ The values for the integral $J^-$ (mb). }
\label{tab:3}       
\begin{center}
\begin{tabular}{ll}
\hline\noalign{\smallskip}
Source & $J^-$ (mb)\\
\noalign{\smallskip}\hline\noalign{\smallskip}
H\"ohler-Kaiser \cite{hohler2} & -1.06 \\
Koch \cite{koch3} & -1.077 $\pm$ 0.047 \\
Gibbs {\it et al.} \cite{gibbs} & -1.051 $\pm$ 0.005$^{\rm a}$\\
Ericson {\it et al.} \cite{elt} & -1.083 $\pm$ 0.032 \\
Present work  & -1.060 $\pm$ 0.030\\
\noalign{\smallskip}\hline
$^{\rm a}$Statistical error only.
\end{tabular}
\end{center}
\end{table}
To assign an error bar to our result for $J^-$, we note that
the statistical uncertainty is quite small, about  0.007 mb. It is, however,
hard to estimate the systematic uncertainty. The systematic normalization
uncertainties of the data have been taken into account in the fitting procedure \cite{pekko}.
Also, there remain discrepancies in the data, see the discussion, {\it e.g.}, in
ref. \cite{hohler2}. The $\Delta$-resonance gives a major contribution to
the $J^-$ -integral and in that range the two accurate experiments, Carter {\it et al.} \cite{carter}
and Pedroni {\it et al.} \cite{pedroni}, differ slightly, but in a systematic manner.
The impact of the difference on the $J^-$ is 0.012 mb. We adopt this as the
number reflecting experimental systematic effects.
Furthermore, 
the forward dispersion relations, which have been imposed as contraints,
do need a value for the pion-nucleon coupling constant as input.
Here the  pion-nucleon coupling constant has been allowed to vary
and the effect on the error bar for the $J^-$ integral is 0.001 mb.
For the asymptotic part, where data do not exist, the average of the numbers in table 1
has been taken and the corresponding error is chosen such that all the displayed values fall
within the errors. This gives as the estimate for the error 0.004 mb.
If in the integrand for $J^-$ involving $\sigma^{\rm Tot}_{\pi^-p}-\sigma^{\rm Tot}_{\pi^+p}$
a Coulomb correction of the type \cite{elt} is adopted in the range 0.725 - 2.03 GeV/c, the value
for $J^-$ changes by 0.6 \%. Adding up all these errors, we obtain the quoted uncertainty 0.030 mb.

\section{Results and discussion}
\label{St}

From eq. (1) the pion-nucleon coupling constant can be extracted with the result
\begin{eqnarray}
f^2&=& \frac{1}{2} [1-(\frac{\mu}{2 m})^2] \times \; \; \; \; \nonumber \\ 
 &  & \; \; \;[\frac{1}{2}(1+\frac{\mu}{m})(a_{\pi^- p} - a_{\pi^+ p})\mu - J^-\mu^2].
\end{eqnarray}
The contribution to $f^2$ from the term involving the difference of
the $\pi^-p$ and $\pi^+p$ scattering lengths is about 2/3 and
the $J^-$ piece about 1/3.
With the scattering lengths and the integral $J^-$ evaluated in sects. 2-4, the 
coupling constant becomes
\begin{eqnarray}
f^2 = 0.075 \pm 0.002.
\end{eqnarray}
This precision can not compete with the expected accuracy from other methods
involving the dispersion relations for the $B$-amplitudes, but we shall use
this range of possible values as input for the phase shift analysis
with fixed-$t$ constraints.

In the discussion above isospin symmetry has been used only in fixing the
threshold values for the total cross sections, a statement valid to
${\cal O}(p^3)$ in chiral perturbation theory. By invoking
the isospin invariance we can relate $a_{\pi^- p} - a_{\pi^+ p}
= 2 \, a^-_{0+}$, where the isovector $s$-wave scattering length  is accessible through 
the atomic width measurement, eq. (11). The numbers for $a^-_{0+}$ quoted in sect. 2,
yield the coupling strength in the range $f^2=0.076-0.077$, {\it i.e.} within the error range 
quoted above. By making use of isospin invariance one can avoid the
question of the consistency of the approach when one is extracting the hadronic $\pi^-p$ 
$s$-wave scattering length in chiral perturbation theory and the $\pi^+p$ scattering
length in a discrete phase shift analysis exploiting the Tromborg
approach for the electromagnetic corrections. We are aware of one low-energy $\pi$N analysis
with a complete treatment of both strong and electromagnetic effects
to third order in chiral perturbation theory, the work of Fettes and Mei\ss ner \cite{fm}.
There, indeed, it is found that nonlinear pion-nucleon-photon coupling terms
generate sizeable effects. However, the comparison in ref. \cite{fm} for the
low-energy $\pi^+p$ $s$-wave scattering with the Karlsruhe results, which also
use Tromborg corrections, shows that the effect is relatively small there.

The potential model \cite{sigg} corrections $\delta_\epsilon = (-2.1 \pm 0.5) \times 10^{-2}$
and $\delta_\Gamma = (-1.3 \pm 0.5) \times 10^{-2}$ would lead to coupling
values in the range $f^2=0.077-0.078$ in agreement with ref. \cite{elt}.

\section{Conclusions}
\label{co}

The GMO sum rule provides a tool to relate experimental information on
pion-nucleon scattering from different sources. Also, it is relatively
straightforward to analyze the error propagation. With the input from 
pionic hydrogen and a discrete phase shift analysis for the $\pi^+p$
interaction we
obtain for the pion-nucleon coupling constant the value $f^2=0.075 \pm 0.002$.
It turns out, that the largest uncertainty 
in determining a value for the pion-nucleon coupling constant is
currently due to the large uncertainty in the correction factor $\delta_\epsilon$
for extracting the the $\pi^- p$ $s$-wave scattering length
from the $\epsilon_{1s}$ hadronic level shift measurement. Assuming isospin
invariance we can make use of the isovector $s$-wave scattering
length $a^-_{0+}$ which will be determined within 1 \% by the PSI group in
the future.

\begin{acknowledgement}
We wish to thank 
A.M. Green for useful comments on the manuscript.
One of us (PM) thanks the Magnus Ehrnrooth Foundation and
the Waldemar von Frenckell Foundation for financial support.
Support from the Academy of Finland and the Russian Academy of
Sciences exchange grant is acknowledged.
This work was supported in part by the EU Contract 
MRTN-CT-2006-035482, FLAVIAnet.
\end{acknowledgement}
%

\end{document}